# Zero-Point Energies, the Uncertainty Principle and Positivity of the Quantum Brownian Density Operator

Allan Tameshtit
Email: allan.tameshtit@utoronto.ca

High temperature and white noise approximations are frequently invoked when deriving the quantum Brownian equation for an oscillator. Even if this white noise approximation is avoided, it is shown that if the zero point energies of the environment are neglected, as they often are, the resultant equation will violate not only the basic tenet of quantum mechanics that requires the density operator to be positive, but also the uncertainty principle. By including the zero-point energies, asymptotic results describing the evolution of the oscillator are obtained that preserve positivity and, therefore, the uncertainty principle.

Quantum Brownian motion arises when a system of interest interacts with its environment, and is therefore ubiquitous in nature. One theoretical approach for studying this phenomenon involves the scattering of particles by an otherwise free, heavier particle [1,2]. Alternatively, in the approach adopted below, a bound oscillator coupled to a heat bath is analyzed, which, by setting the appropriate frequency to zero, can include a free particle coupled to the bath as a limiting case. Efforts to derive from first principles a quantum Brownian equation (QBE) for such an oscillator have been extensive ([3-5] and references therein). A major obstacle that has hindered attempts in this direction is the basic tenet of quantum mechanics that requires any density operator obeying the QBE to remain positive [6]. After a temperature expansion, a positivity-preserving equation was obtained in Ref.[7] by neglecting terms that are no smaller than some appearing in the final equation. However, this equation is not valid at the small times where positivity failure originates. The results herein establish that rather than a low-order temperature expansion, it is inclusion of the zero-point energies of the environment that is pivotal.

From what is perhaps the most popular model for studying quantum Brownian motion, a harmonic oscillator coupled to a heat bath with a factorized initial condition, an *exact* QBE has been derived in the literature [3] describing the evolution of the oscillator's density operator, $\rho$:

$$\frac{d\rho}{dt} = \frac{1}{\hbar i}\left[\frac{p^2}{2m} - \frac{m}{2}f_{pq}(t)q^2, \rho(t)\right] - \frac{1}{\hbar i}f_{pp}(t)\left[q, \frac{1}{2}(p\rho + \rho p)\right]$$
$$+ \frac{1}{\hbar^2}d_{pq}(t)[p,[q,\rho(t)]] - \frac{m}{\hbar^2}d_{pp}(t)[q,[q,\rho(t)]].$$
(1)

Sparing the reader the rather unwieldy expressions for the coefficients $f_{pq}$, $f_{pp}$, $d_{pp}$ and $d_{pq}$, it will suffice to mention that these coefficients depend not only on time, but some also on temperature via frequency integrals that involve the bath energy equilibrium values, $\hbar\omega\left(\bar{n} + 1/2\right)$, where $\bar{n}$ are the average occupation numbers $\left(e^{\hbar\omega/kT} - 1\right)^{-1}$. It is common practice to invoke the high-temperature approximation

$$\hbar\omega\left(\bar{n}+1/2\right)\to kT \tag{2}$$

followed by a white noise approximation to render the foregoing coefficients time independent, resulting in an autonomous equation that is analogous to the classical Brownian equation [3-5]:

$$\frac{d\rho}{dt}=\frac{1}{\hbar i}\left[\frac{p^2}{2m}+\frac{m}{2}\Omega_1^2 q^2,\rho(t)\right]+\frac{\Gamma}{\hbar i}[q,p\rho+\rho p]-\frac{2\Gamma kTm}{\hbar^2}[q,[q,\rho(t)]], \tag{3}$$

where the constant $\Gamma$ is a measure of the coupling strength between the harmonic oscillator and the bath, and the constant $\Omega_1$ is the shifted harmonic oscillator frequency. Though this last equation is simpler than Eq.(1), the simplifying approximations leading thereto come at a price. Because the last two terms of Eq.(3) are not expressible in Lindblad form (i.e., cannot be re-written as a sum $-\sum_k \{A_k,\rho,A_k\}$ where $\{A,\rho,B\}\equiv BA^\dagger\rho+\rho BA^\dagger-2A^\dagger\rho B$) [8], Eq.(3) does not generally preserve positivity of the density operator [3, 9-11]. On the other hand, it is interesting to note that although the last three terms of Eq.(1) are also not expressible in Lindblad form, Eq.(1), being exact, does preserve positivity, managing to achieve this feat because its coefficients depend on time.

This suggests that a simpler, positivity-preserving equation might be obtained from Eq.(1) by making the usual high temperature approximation (2), but abandoning the white noise approximation that leads to the autonomous Eq.(3).

Unfortunately, herein it is shown that the QBE so obtained does not preserve positivity. To repeat, even if the white noise approximation is avoided to allow the QBE coefficients to depend on time, the popular high temperature approximation (2) damns the resultant density operator to be generally non-positive. It is further shown in the appropriate regime that if instead of the high temperature approximation (2) the more uniform approximation

$$\hbar\omega\left(\bar{n}+1/2\right)\to kT+\hbar\omega/2 \tag{4}$$

is adopted, the resultant non-autonomous equation *does* preserve positivity. Thus, inclusion of the zero-point bath energies, the "$\hbar\omega/2$" in the expression $\hbar\omega\left(\bar{n}+1/2\right)$, appears to be crucial.

Our starting point is the exact solution of Eq.(1) in operator form [12],

$$\rho(t) = \exp\left[\frac{\ln R^2}{2\hbar^2(1-R^2)}\left(\begin{array}{c}mY\{q,\cdot,q\}+\dfrac{1}{m}X\{p,\cdot,p\}\\-\dfrac{1}{2}\left(\dot{X}-i\hbar(1-R^2)\right)\{p,\cdot,q\}-\dfrac{1}{2}\left(\dot{X}+i\hbar(1-R^2)\right)\{q,\cdot,p\}\end{array}\right)\right]$$
$$\times N(t)\tilde{M}(t)\rho(0)\tilde{M}(t)^\dagger \qquad N(t)^\dagger,$$
(5)

where the following definitions are similar to those found in the Wigner function solution in [3] and arise with the use of Ullersma's [13] spectral strength:

(i) $X(t) = \dfrac{2\kappa\alpha^2\hbar}{\pi}\int_0^\infty d\omega \dfrac{\omega}{\alpha^2+\omega^2}\left|\int_0^t dt' e^{i\omega t'}A(t')\right|^2 \left(\bar{n}+1/2\right)$ (6)

with

$$A(t) = \frac{2\Gamma\left[e^{-(\alpha-2\Gamma)t}-e^{-\Gamma t}\cos(\Omega t)\right]+\Omega^{-1}\left[(\alpha-2\Gamma)^2+\Omega^2-\Gamma^2\right]e^{-\Gamma t}\sin(\Omega t)}{(\alpha-3\Gamma)^2+\Omega^2},$$
(7)

where $\alpha$ plays the role of a high-frequency cut-off of the bath, $\kappa = 2\Gamma\left((\alpha-\Gamma)^2-\Omega^2\right)/\alpha^2$ and $\Omega$ is approximately equal to the shifted frequency $\Omega_1$ when $\Gamma$ is small compared to $\Omega_1$ and $\alpha$;

(ii) $Y(t)$ is the same expression as $X(t)$ but with $A$ replaced by $dA/dt'$; and

(iii) $R(t) = \sqrt{\left(\dfrac{dA}{dt}\right)^2 - A\dfrac{d^2A}{dt^2}}$, where to ensure the radicand is positive, it is assumed that $\alpha \geq 3\Gamma$.

The explicit expression for the unitary operator $N(t)\tilde{M}(t)$, which satisfies $N(0)\tilde{M}(0) = 1$, is given in [12], but need not concern us here.

The following theorem will be used.

*Theorem*
Suppose an allowable (i.e., normalized, self-adjoint and positive) initial state $\rho(0)$ evolves according to

$$\rho(t) = \exp\left[\frac{\ln(1-r^2)}{2r^2\hbar^2}\left(ma\{q,\cdot,q\}+\frac{b}{m}\{p,\cdot,p\}-\frac{1}{2}(c-i\hbar r^2)\{p,\cdot,q\}-\frac{1}{2}(c+i\hbar r^2)\{q,\cdot,p\}\right)\right]$$
$$\times V(t)\rho(0)V(t)^\dagger \qquad (8)$$

where $a \geq 0, b \geq 0, 0 \leq r < 1$, and $c$ are real parameters that depend on $t$ subject to $a(0) = b(0) = c(0) = r(0) = 0$, and $V(t)$ is a unitary operator satisfying $V(0) = 1$. Letting the symbol $\langle \ \rangle_V$ denote expectation values with respect to the state $V(t)\rho(0)V(t)^\dagger$ (e.g., $\langle qp + pq \rangle_V \equiv \text{Tr}(qp + pq)V(t)\rho(0)V(t)^\dagger$), suppose further that at some time $t' > 0$ the following four conditions involving the initial state $\rho(0)$ are obeyed:

$$d_1 \equiv \frac{1}{4}\left(\hbar^2 r^2 + c\langle qp + pq\rangle_V\right)^2 - ab\left(\hbar^2 + \langle qp + pq\rangle_V^2\right) > 0 \tag{9}$$

$$\hbar^2 r^2 + c\langle qp + pq\rangle_V > 0 \tag{10}$$

$$\langle q^2\rangle_V \langle p^2\rangle_V = \frac{1}{4}\left(\langle qp + pq\rangle_V^2 + \hbar^2\right), \text{ and} \tag{11}$$

$$\langle q^2\rangle_V \in (s_-, s_+) \tag{12}$$

where

$$s_\pm \equiv \frac{1}{2ma}\left(\frac{1}{2}(\hbar^2 r^2 + c\langle qp + pq\rangle_V) \pm d_1^{1/2}\right) \text{ with } a \neq 0. \tag{13}$$

Then $\rho(t')$ is non-positive provided that at time $t'$

$$0 \leq 4ab - c^2 < \hbar^2 r^4. \tag{14}$$

*Proof:*
Consider the quadratic form in $x$,

$$amx^2 - \frac{1}{2}\left(\hbar^2 r^2 + c\langle qp + pq\rangle_V\right)x + \frac{b}{4m}\left(\hbar^2 + \langle qp + pq\rangle_V^2\right), \tag{15}$$

where this and all other expressions in this proof are evaluated at $t'$ unless otherwise indicated. Its discriminant is $d_1$. Because of inequality (9), expression (15) has a pair of real, distinct roots $s_\pm$, and because of inequality (10) and the assumptions that $a > 0$ and $b \geq 0$, the smaller root $s_-$ is non-negative and the larger root $s_+$ is positive. Consequently, on account of assumption (12),

$$am\langle q^2\rangle_V^2 - \frac{1}{2}\left(\hbar^2 r^2 + c\langle qp + pq\rangle_V\right)\langle q^2\rangle_V + \frac{b}{4m}\left(\hbar^2 + \langle qp + pq\rangle_V^2\right) < 0. \tag{16}$$

Using Eq.(11), inequality (16) may be re-written as

$$\hbar^2 - 4(1 - r^2)^2 \langle q^2\rangle_V \langle p^2\rangle_V - 4ma(1 - r^2)\langle q^2\rangle_V - 4(1 - r^2)\frac{b}{m}\langle p^2\rangle_V - \hbar^2 r^4 \tag{17}$$

$$+ (1 - r^2)^2\langle qp + pq\rangle_V^2 + 2(1 - r^2)c\langle qp + pq\rangle_V > 0.$$

Now introduce a parameter

$$\lambda = \frac{\hbar + w^{1/2}}{2\left((1 - r^2)\langle p^2\rangle_V + ma\right)} \tag{18}$$

where $w$ is defined to be the left hand side of inequality (17). By virtue of inequality (17), $\lambda$ is real, and by construction, $\lambda$ is a root that satisfies

$$\left((1-r^2)\langle p^2\rangle_V + ma\right)\lambda^2 - \hbar\lambda + (1-r^2)\langle q^2\rangle_V + \frac{b}{m}$$

$$-\frac{ab - \frac{\hbar^2}{4}r^4 + \frac{1}{4}(1-r^2)^2\langle qp+pq\rangle_V^2 + \frac{1}{2}(1-r^2)c\langle qp+pq\rangle_V}{(1-r^2)\langle p^2\rangle_V + ma} = 0. \quad (19)$$

In another vein, it may be noted [14] that the evolution governed by Eq.(8) yields the following expectation values for all $t \geq 0$:

$$\langle q^2\rangle = (1-r^2)\langle q^2\rangle_V + \frac{b}{m} \quad (20)$$

$$\langle p^2\rangle = (1-r^2)\langle p^2\rangle_V + ma \quad (21)$$

$$\langle qp+pq\rangle = (1-r^2)\langle qp+pq\rangle_V + c. \quad (22)$$

These last three expressions allow us to write

$$I(\lambda,\beta;t) \equiv \mathrm{Tr}(q+(\beta+i\lambda)p)\rho(t)(q+(\beta-i\lambda)p) = \left((1-r^2)\langle p^2\rangle_V + ma\right)\beta^2$$

$$+\left((1-r^2)\langle qp+pq\rangle_V + c\right)\beta + \lambda^2\left((1-r^2)\langle p^2\rangle_V + ma\right) + (1-r^2)\langle q^2\rangle_V + \frac{b}{m} - \hbar\lambda \quad (23)$$

where $\beta$ is a real parameter. Considering the right hand side of Eq.(23) as a quadratic function in $\beta$, we conclude that $I(\lambda,\bar{\beta};t') < 0$ provided the discriminant satisfies

$$\left((1-r^2)\langle qp+pq\rangle_V + c\right)^2$$

$$-4\left((1-r^2)\langle p^2\rangle_V + ma\right)\left(\lambda^2\left((1-r^2)\langle p^2\rangle_V + ma\right) + (1-r^2)\langle q^2\rangle_V + \frac{b}{m} - \hbar\lambda\right) > 0 \quad (24)$$

and provided $\bar{\beta}$ lies between the two roots of the right hand side of Eq.(23) evaluated at $t'$, which are both real when inequality (24) is imposed. It is straightforward to show that when Eq. (19) holds, inequality (24) is equivalent to the simpler inequality $4ab - c^2 - \hbar^2 r^4 < 0$, which is part of assumption (14). We have thus shown that with the foregoing assumptions, $I(\lambda,\beta;t')$ is negative when $\lambda$ and $\beta$ are given by expression (18) and $\bar{\beta}$, respectively. This implies $\rho(t')$ is non-positive, which completes the proof.

If in addition to the hypotheses of the theorem, we also assume $\langle q\rangle(t') = \langle p\rangle(t') = 0$, then from expressions (18), (20)-(22) and (24) we obtain at time $t'$

$$(\Delta q)^2(\Delta p)^2 < \left(\langle \tilde{q}\tilde{p}+\tilde{p}\tilde{q}\rangle^2 + \hbar^2 - w\right)/4 \quad (25)$$

where $\tilde{q} \equiv q - \langle q \rangle$ and $\tilde{p} \equiv p - \langle p \rangle$, in violation of the uncertainty principle [15]. We further note that some of the hypotheses of the theorem can be redundant [17].

The evolution Eq.(5) is of the form of Eq.(8) if we associate $a$ with $Y$, $b$ with $X$, $c$ with $\dot{X}$, $V$ with $N\tilde{M}$, and $r^2$ with $1-R^2$. We assume that each of $\alpha, \kappa$ and $t$ is positive. We check that assumptions (9)-(14) are true for the oscillator at some time $t' > 0$, which, as will now be shown, can be short. When the high temperature approximation (2) is employed, the following asymptotic results are obtained as $t \to 0$.

$$X = \frac{kTt^4}{4}\kappa\alpha + o(t^4) \tag{26}$$

$$\frac{dX}{dt} = kTt^3\kappa\alpha + o(t^3) \tag{27}$$

$$Y = kTt^2\kappa\alpha + o(t^2) \tag{28}$$

and

$$1 - R^2 = \alpha^2 t^3 \kappa/6 + o(t^3). \tag{29}$$

When these expressions are inserted into definition (9), we find that there exists a sufficiently small time at which the discriminant $d_1$ is greater than zero if we insist that at that time

$$\langle qp + pq \rangle_V = 3\frac{kT}{\alpha}. \tag{30}$$

With reference to assumption (14), we note that $0 \leq 4ab - c^2$ holds for all $t \geq 0$ under any positive approximation of $\bar{n}+1/2$, which follows from the definitions of $X$, $Y$ and the Cauchy-Schwartz inequality for integrals. In addition,

$$4ab - c^2 - \hbar^2 r^4 = -\left(\frac{\hbar\alpha^2 t^3 \kappa}{6}\right)^2 + o(t^6), \tag{31}$$

which is less than zero if $t$ is small enough. Given Eq.(30), assumption (10) also follows for small enough $t$. Hence, $\exists t' > 0$ such that by choosing the initial state

$$\rho(0) = M^\dagger(t')N^\dagger(t')\chi N(t')\tilde{M}(t'), \tag{32}$$

with $\chi$ being the pure density operator corresponding to the Wigner function

$$\frac{1}{\pi\hbar}\exp\left[-\frac{2}{\hbar^2}\left(\frac{\hbar^2 + (3kT/\alpha)^2}{4\bar{s}}q^2 + \bar{s}\,p^2 - 3\frac{kT}{\alpha}pq\right)\right] \tag{33}$$

with $\bar{s} \equiv [s_+(t') + s_-(t')]/2$, assumptions (9)-(14) are true. We thus conclude that making the oft-used high-temperature approximation (2) yields evolution that violates the positivity requirement and, in view of expression (25) and the fact that $\langle q \rangle(t') = \langle p \rangle(t') = 0$ when the initial state is Eq.(32), violates the uncertainty principle even if the coefficients in the QBE are allowed to depend on time [18].

Given that the only approximation in play is expression (2), and that this approximation would appear to be innocuous at high temperatures, why is expression (2) not good enough to preserve positivity, no matter how high the temperature? A partial answer is that expressions (26)-(29) yield $4XY - \dot{X}^2 = o(t^6)$ and $\hbar^2(1-R^2)^2 \sim (\hbar\alpha^2\kappa)^2 t^6/36$; thus, the latter dissipation contribution is larger than the former fluctuation contribution at small times no matter how high the temperature, and the left hand side of (31) is consequently negative. Moreover, the use of the continuous, as opposed to a discrete, spectral strength in deriving Eq. (1) gives rise to larger values of the dissipation factor $(1-R^2)^2$ at small times.

In contrast, if instead of the high temperature approximation (2) we were to use the more uniform expression (4), then positivity is preserved at small times [19]. To demonstrate this, we first need the following corollary to Theorem 2 of Ref. [8] (cf., also, Eq.(27) of Ref.[12]).

*Corollary*

Let $\sigma$, $\eta$ and $\xi$ be real, time-dependent parameters, and $\zeta$ a time-dependent complex parameter, and let $\rho(0)$ be an allowable initial state. Then the density operator

$$\exp\left[-\sigma(t)\left(\eta(t)\{q,\cdot,q\} + \xi(t)\{p,\cdot,p\} + \zeta(t)\{q,\cdot,p\} + \zeta^*(t)\{p,\cdot,q\}\right)\right]\rho(0) \quad (34)$$

is positive for any $t \geq 0$ at which $\sigma \geq 0$, $\eta \geq 0$, $\xi \geq 0$ and $\eta\xi \geq |\zeta|^2$.

*Proof*:

As can be shown using the Cauchy-Schwartz inequality and as has already been pointed out in [20], the sum of operators in the exponent in expression (34) may be re-written in Lindblad form, i.e.,

$$-\left(\eta(t)\{q,\cdot,q\} + \xi(t)\{p,\cdot,p\} + \zeta(t)\{q,\cdot,p\} + \zeta^*(t)\{p,\cdot,q\}\right)$$
$$= -\sum_{n=1} \{a_n(t)q + b_n(t)p,\cdot,a_n(t)q + b_n(t)p\}, \quad (35)$$

for some generally complex numbers $a_n$ and $b_n$, provided $\eta \geq 0$, $\xi \geq 0$ and $\eta\xi \geq |\zeta|^2$. Thus, if these last three inequalities are true,

$$L \equiv -\sigma(t)\left(\eta(t)\{q,\cdot,q\} + \xi(t)\{p,\cdot,p\} + \zeta(t)\{q,\cdot,p\} + \zeta^*(t)\{p,\cdot,q\}\right) \quad (36)$$

may also be re-written in Lindblad form:

$$L = -\sum_n \{\sigma^{1/2}(t)a_n(t)q + \sigma^{1/2}(t)b_n(t)p,\cdot,\sigma^{1/2}(t)a_n(t)q + \sigma^{1/2}(t)b_n(t)p\}. \quad (37)$$

It is implicit in [8] (assuming the results therein also apply to the unbounded operators in Eq. (37)) that the solution of the differential equation $d\rho/d\tau = L\rho$, with initial condition $\rho(0)$, is positive for all $\tau \geq 0$ since $L$ is of Lindblad form. In this last differential equation, $\tau$ is considered to be the "time" and $t$ a parameter. But expression (34) is the solution of this last differential equation at $\tau = 1$. Hence, the operator (34) is positive for any $t \geq 0$ at which $\sigma \geq 0$, $\eta \geq 0$, $\xi \geq 0$ and $\eta\xi \geq |\zeta|^2$ if $\rho(0)$ is an allowable initial state, which proves the corollary to Theorem 2 of Ref.[8].

The evolution Eq.(5) is in the form of Eq.(34) if we associate $\sigma$ with $-\dfrac{\ln R^2}{2\hbar^2(1-R^2)}$, $\eta$ with $mY$, $\xi$ with $X/m$ and $\zeta$ with $-\left(\dot{X}+i\hbar(1-R^2)\right)/2$. Using approximation (4) instead of (2), the following asymptotic results are obtained as $t \to 0$.

$$X = \frac{\kappa\alpha^2\hbar}{4\pi}\left(\pi\frac{kT}{\hbar\alpha}+\frac{7}{4}-\gamma-\ln\alpha t\right)t^4 + o(t^4) \tag{38}$$

$$\frac{dX}{dt} = \frac{\kappa\alpha^2\hbar}{\pi}\left(\pi\frac{kT}{\hbar\alpha}+\frac{3}{2}-\gamma-\ln\alpha t\right)t^3 + o(t^3), \tag{39}$$

and

$$Y = \frac{\kappa\hbar\alpha^2}{\pi}\left(\pi\frac{kT}{\hbar\alpha}+\frac{3}{2}-\gamma-\ln\alpha t\right)t^2 + o(t^2) \tag{40}$$

where $\gamma$ is Euler's constant 0.577… The quantity $R$, being independent of temperature, continues to satisfy relation (29) under approximation (4). With the use of relations (29) and (38)-(40), it is straightforward to check that $\sigma \geq 0$, $\xi \geq 0$, $\eta \geq 0$ and $\eta\xi \geq |\zeta|^2$ for small enough $t$. Thus, by the foregoing corollary, we see that the approximation (4) ensures that positivity is preserved for small times. The author has elsewhere shown [21] that for large values of $\alpha/\Omega$, $\alpha/\Gamma$ and $kT/\hbar\Omega$, with $\alpha\Gamma \leq \Omega^2/2$, the approximation (4) leads to positivity for *all* $t \geq 0$.

The greater-than-zero energy lower bound of an oscillator—its zero point energy—prevents its position and momentum from being overly determined, sparing the uncertainty principle. The results herein show that the zero point energies of the *environment* also play a key role: if the high temperature approximation (2) that ignores these zero point energies is invoked, then not only is the positivity requirement violated, but so too is the uncertainty principle; moreover, in the appropriate regime, if approximation (4) that includes these energies is used then the density operator remains positive.

Eqs.(5, 38-40), preserving positivity, afford the opportunity of investigating entanglement decoherence that arises when the initial factorized bath-system state rapidly entangles in an inner limit on the order of $1/\alpha$ as the total system seeks local equilibrium. It is hoped that either numerical calculations [22] or increasingly fast experimental probes might be able to explore this regime. At a theoretical level, the inner limit was studied in Ref.[12], but because dissipation was neglected in the inner limit, the zero-point energies were not confronted there. Extension of the work in Ref.[12] to include this dissipation and the zero point energies of the environment will be left for a future communication.